\documentclass[fleqn,12pt,twoside]{article}
\usepackage[headings]{espcrc1}

\readRCS
$Id: espcrc1.tex,v 1.2 2004/02/24 11:22:11 spepping Exp $
\usepackage{graphicx}

\title{Astrophysical Rates for Explosive Nucleosynthesis: Stellar and Laboratory Rates for Exotic Nuclei}

\author{T. Rauscher\address{Department of Physics, University of Basel,\\
        Klingelbergstr. 82, 4056 Basel, Switzerland}%
        \thanks{Supported by the Swiss NSF, grant 200020-122287.}}
       
\runtitle{Astrophysical Rates for Explosive Nucleosynthesis}

\begin{document}

\maketitle

\begin{abstract}
A selected overview of stellar effects and reaction mechanisms with relevance to the prediction of astrophysical reaction rates far off stability is provided.
\end{abstract}

\section{INTRODUCTION}

Nuclear theory is important in the determination of reaction cross
sections and rates for astrophysics in several respects. Firstly, a large
number of reactions in astrophysical environments involve highly unstable
nuclei which are unaccessible in laboratory measurements. This is
especially true for the high temperature plasmas of stellar explosions.
Secondly, despite of high plasma temperatures, the relevant particle interaction energy
is low by nuclear physics standards. This is a challenge for experimentalists, especially when
studying charged-particle reactions at or below the Coulomb barrier. Very small
cross sections at low energy may even prevent a measurement and have to be
predicted in this case.
Thirdly, the nuclei occur in excited states because they are in thermal
equilibrium with the stellar plasma. This modifies the reaction cross
sections and has important consequences for the determination of stellar
reaction rates. Laboratory measurements only account for a fraction
of the possible transitions and the actual plasma corrections (stellar
enhancement factors, SEF) can only be provided by theory. (Electron screening
is also important but will not be discussed here.)

Large-scale predictions of reaction cross sections across the chart
of nuclei are required to provide the rates needed for reaction networks
applied to nucleosynthesis. In astrophysical applications usually
different aspects are emphasized than in pure nuclear physics investigations.
Reliable predictions of the nuclear properties and optical model ingredients,
like optical potentials for particle and $\alpha$ transmission coefficients \cite{kissprl,rausupp,yalcin,Moalpha,silver,raubranch},
nuclear level densities (NLD) \cite{rau97,mocelj}, resonance energies, and $\gamma$-transition strengths \cite{rau08,litvi},
are needed far from stability. Even at stability, there are still considerable uncertainties, especially in the optical potential for charged particles which is not well constrained at low energy. In addition to the
well-known difficulty in predicting $\alpha$-potentials \cite{yalcin,mohr,somor,koehl,avri}, it was recently
found that also the optical potential for protons may require special attention \cite{kissprl,rausupp,kissold,raujpconf}.

\section{STELLAR ENHANCEMENT}

The SEFs, given by the ratio of stellar rate to ground state rate $r^*/r_\mathrm{g.s.}$ and being a result of thermal population of nuclear excited states, are mainly important at the comparatively high temperatures of explosive burning ($r-$, $p-$, $rp-$, $\nu p$-processes),
although they can play a role in the $s-$process in some cases, when low-lying (a few keV excitation energy) states are present. Because the number of possible transitions to the more strongly bound nucleus is larger, it is generally assumed that stellar effects are less
pronounced in the direction of positive $Q-$value and measurements
should preferrably be performed in this reaction direction to obtain rates closer to the stellar value. However,
recently we showed that the general $Q-$value rule does not apply
for a number of cases due to the suppression of low-energy transitions
in the exit channel by an additional barrier (Coulomb or centrifugal)
\cite{kissprl,rausupp}. effectively reducing the number of contributing transitions.  The
application of detailed balance implies that the populations of target and final nucleus are in equilibrium. It has to be noted, however, that the validity of detailed
balance cannot be taken for granted in all cases. This is well known
for nuclei having isomeric states but may also become
problematic in nuclei with low NLD or for multi-particle
emission channels.

\section{$\gamma$-RAY ENERGIES}

\begin{figure}
\begin{minipage}{14pc}
\includegraphics[width=14pc,angle=-90,clip]{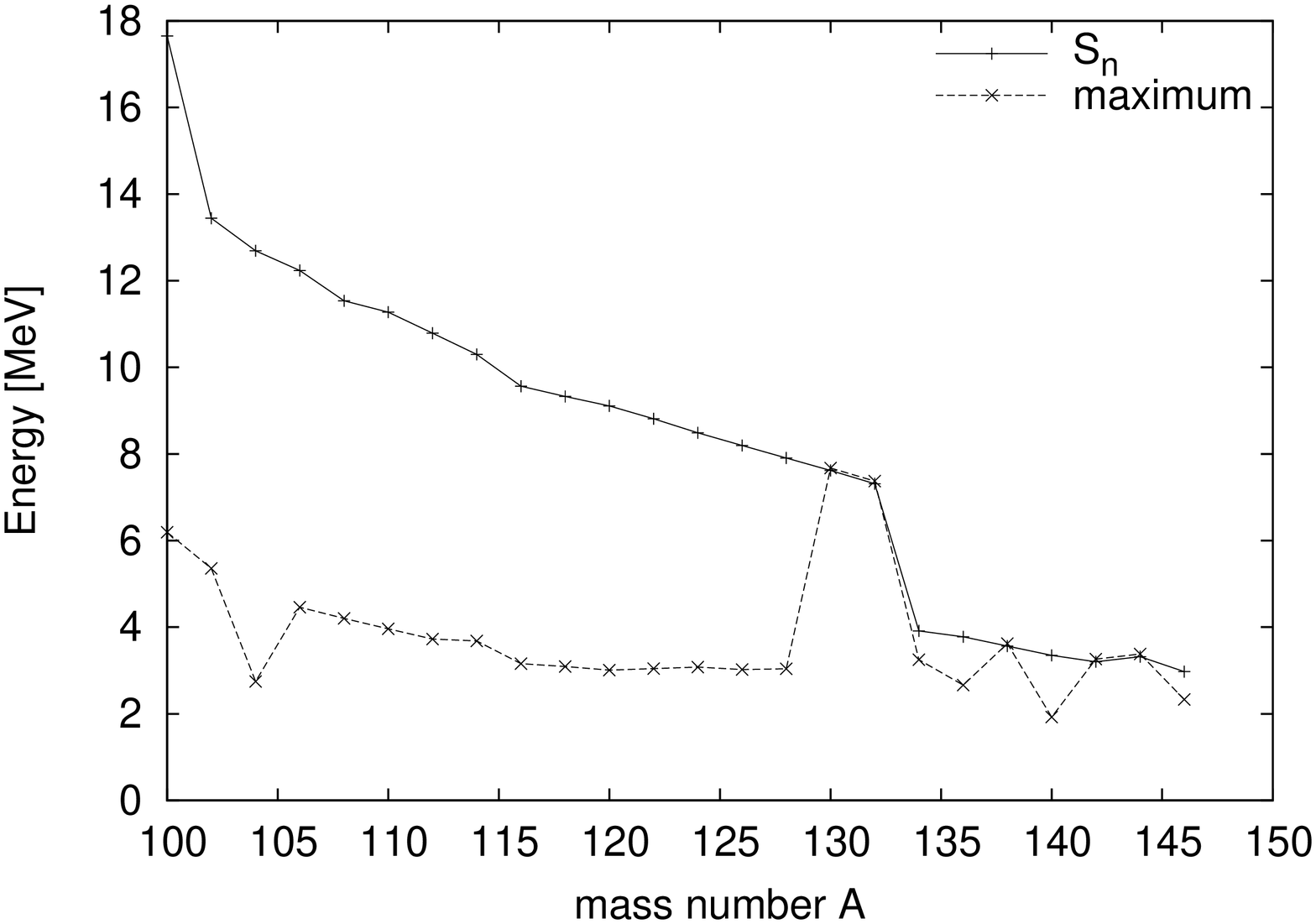}
\end{minipage}\hspace{5.5pc}%
\begin{minipage}{14pc}
\includegraphics[width=14pc,angle=-90,bb=50 100 554 770,clip=true]{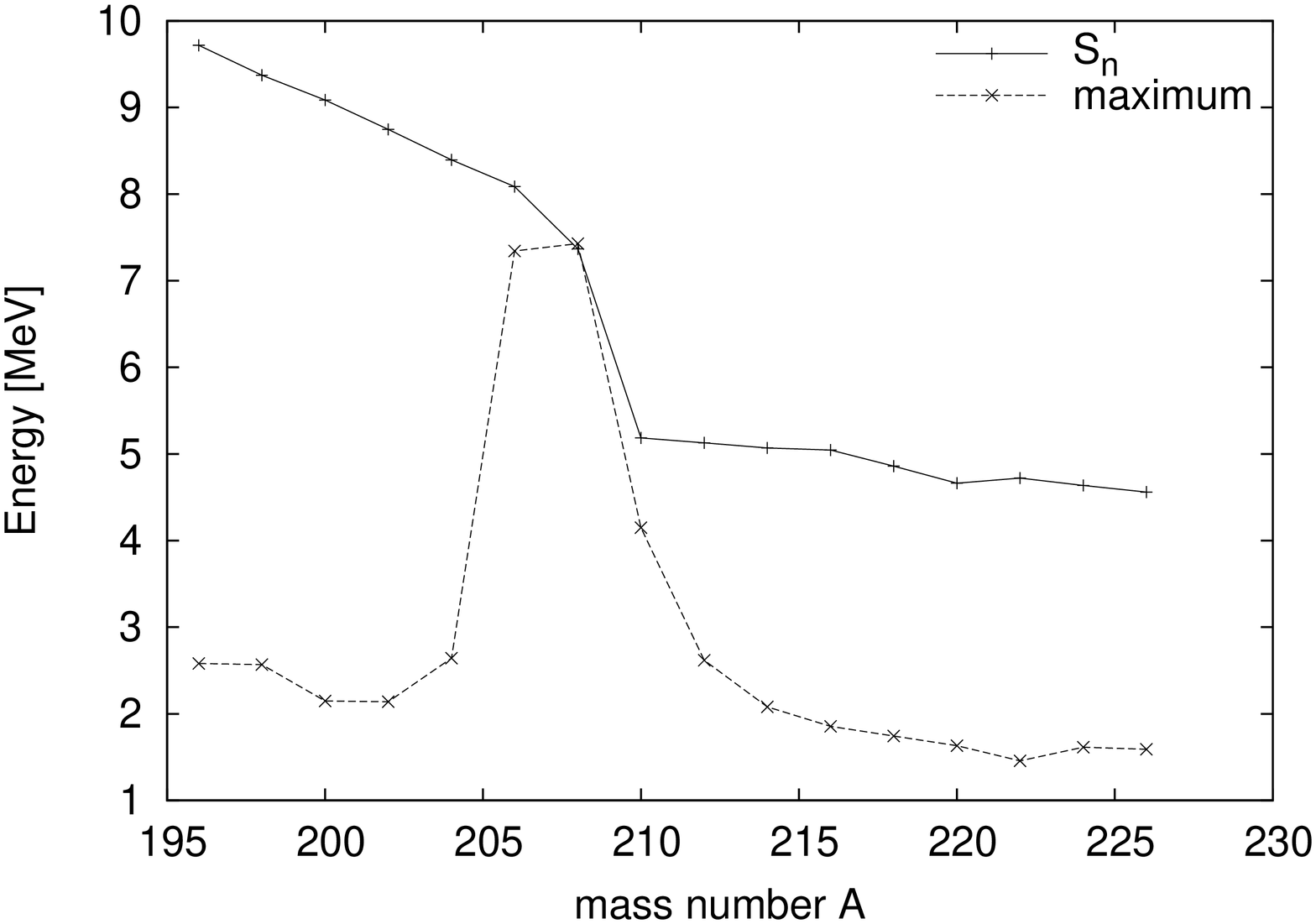}
\end{minipage} 
\caption{\label{fig:gamma}The maximally contributing $\gamma$ energies when capturing 60 keV neutrons on isotopes of Sn (left) and Pb (right) are compared to the neutron separation energies $S_\mathrm{n}$ in the compound nuclei \cite{rau08}. The mass number $A$ is the one of the final (compound) nucleus.}
\end{figure}
In particle capture reactions, $\gamma$ rays with energies in the range $0 < E_\gamma \leq S_\mathrm{particle}+E_\mathrm{particle}$ can be emitted. Due to the fact that the number of accessible levels is increasing exponentially with excitation energy $E_\mathrm{x}$ but the $\gamma$ strength is decreasing with $E_\mathrm{x}$ (because $E_\gamma=S_\mathrm{particle}+E_\mathrm{particle}-E_\mathrm{x}$), only a limited range of $E_\gamma$ will significantly contribute to the cross section, described by a peak around an energy of highest impact $E_\mathrm{sig}$ \cite{rau08}. This energy $E_\mathrm{sig}$ is shown in Fig.\ \ref{fig:gamma} for sequences of Sn and Pb isotopes. It is interesting to note that it remains mostly in the interval of $2-4$ MeV despite large differences in the neutron separation energy $S_\mathrm{n}$. (This is for 60 keV neutrons; $E_\mathrm{sig} \simeq (2-4) + E_\mathrm{n}$ MeV \cite{litvi}.) The exceptions are nuclei with very low NLD, where the transition to the ground state dominates and $E_\mathrm{sig}=S_\mathrm{particle}+E_\mathrm{particle}$. (Note, however, that in this case direct capture may be dominating; see below). For further details see \cite{rau08}.

\section{DIRECT CAPTURE}

\begin{figure}
\begin{minipage}{14pc}
\includegraphics[width=14pc,angle=-90,clip]{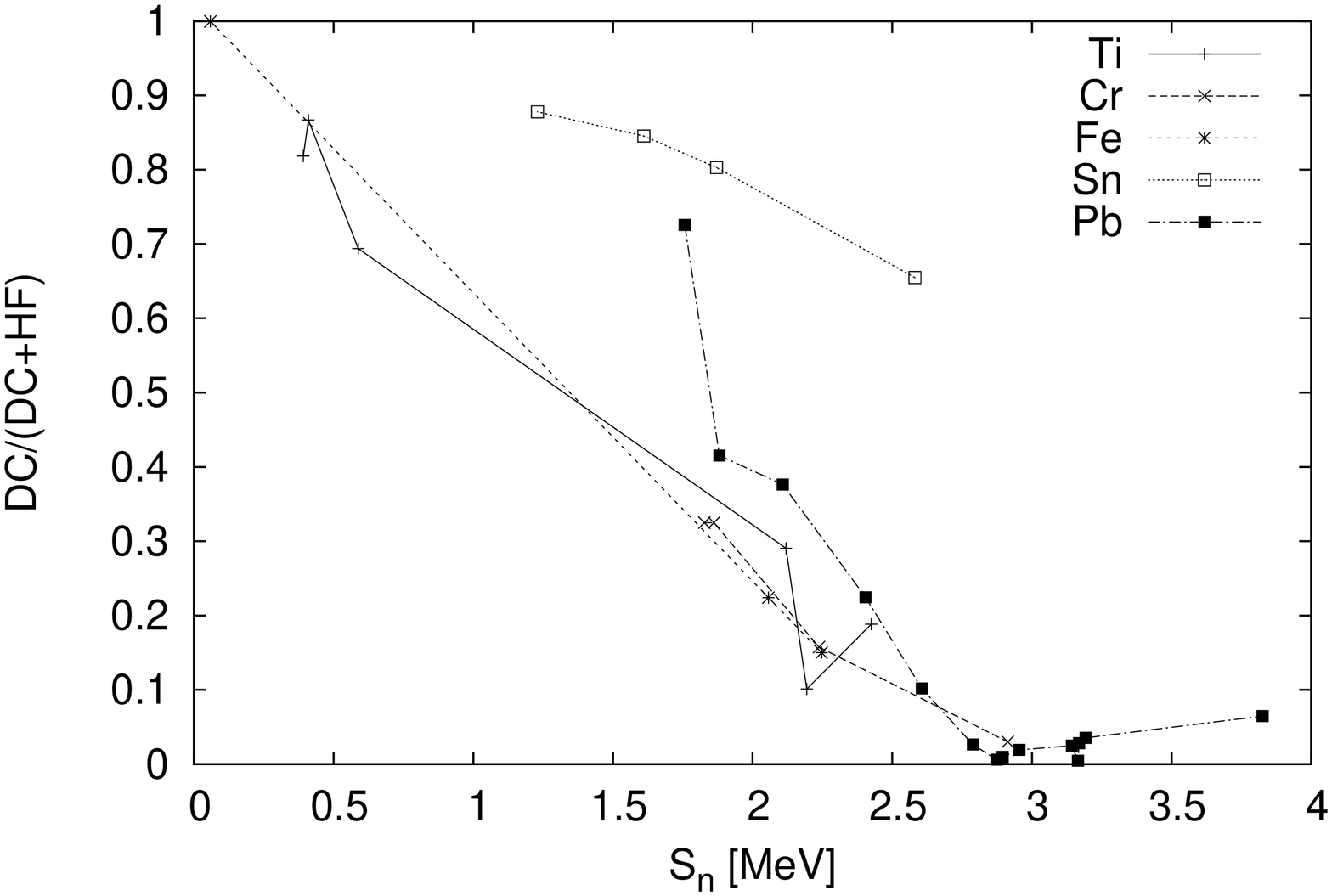}
\end{minipage}\hspace{5.5pc}%
\begin{minipage}{14pc}
\includegraphics[width=14pc,angle=-90,clip]{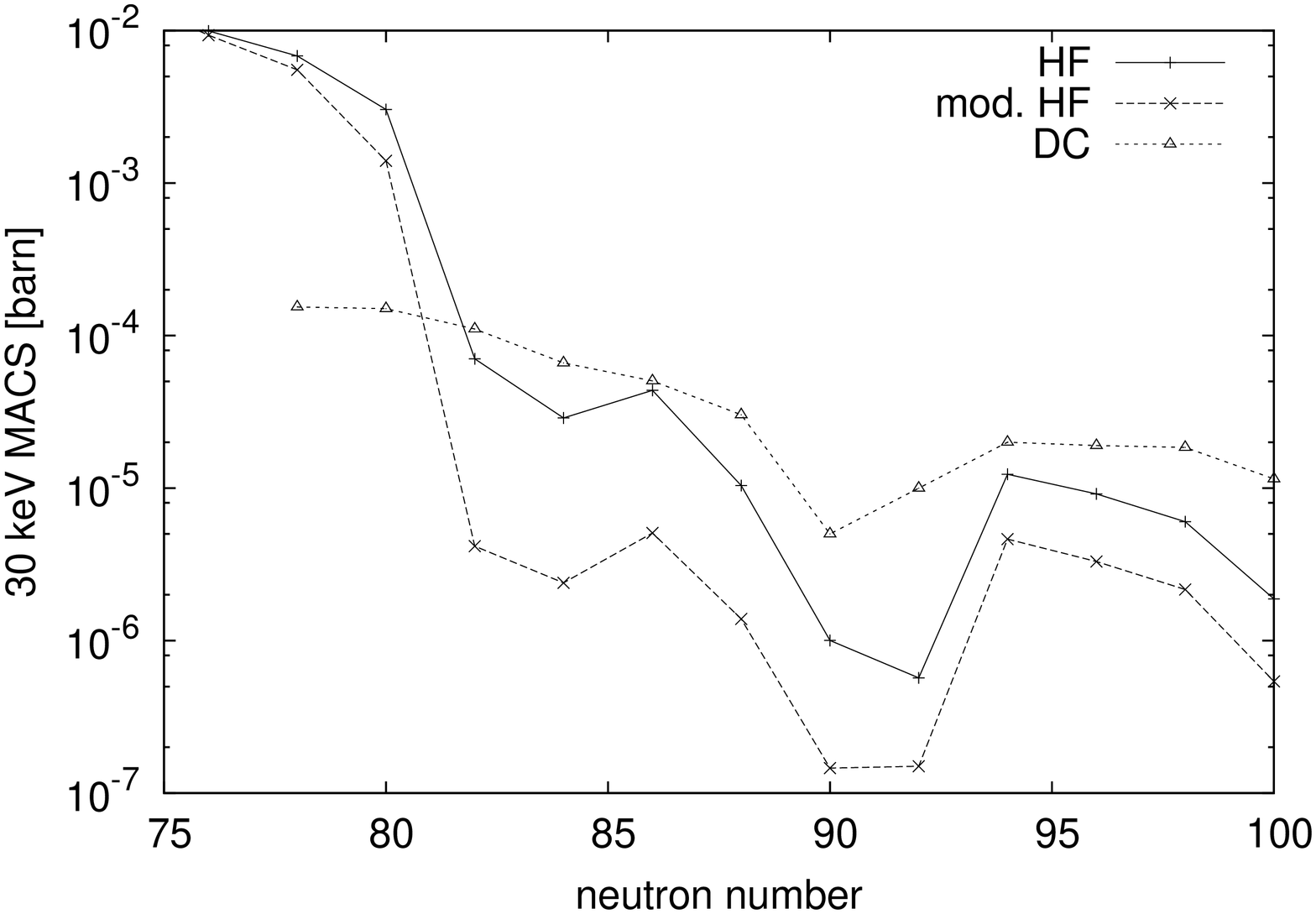}
\end{minipage} 
\caption{\label{fig:dc}Left: Relation between direct neutron capture and (unmodified) compound capture as function of neutron separation energy \cite{raujpg}. Right: Comparison between direct capture with energy-dependent spectroscopic factor, modified Hauser-Feshbach capture, and standard Hauser-Feshbach capture on even Sn isotopes (preliminary results).}
\end{figure}
Most astrophysically relevant reactions can be described in the Hauser-Feshbach
statistical model (HF). Due to the low particle separation energies and/or
NLDs encountered in very neutron- (or proton-) rich nuclei,
direct reactions will also become important, even at very low energies
(in the keV region) \cite{rau97,mocelj,chloup,data,rau00,descrau,rau98,bon07,raujpg}.
Figure \ref{fig:dc} (left) compares direct capture (DC) with the regular HF \cite{raujpg}. It is clearly seen that the relative importance of DC increases with decreasing $S_\mathrm{n}$. This is because the number of final levels for $\gamma$ transitions from HF capture becomes lower. The Sn isotopes have the lowest NLD of the isotopes shown here and consequently the highest relative importance of DC. This comparison was made by simply comparing the DC and HF 30 keV Maxwellian Averaged Cross Section (MACS). Regular HF calculations assume a compound formation probability independent of the NLD at the compound formation energy. The availability of compound states and doorway states defines the applicability of the HF model \cite{rau97}. Relying on an average over resonances the HF model is not applicable with a low NLD at compound formation. On average it will then overpredict the resonant cross section (unless single resonances dominate) because it will overestimate the compound formation probability. This can be treated by introducing a modification of the formation cross section which includes the NLD dependence. For the parity dependence, this was discussed in \cite{loens}. A similar modification but also including the full NLD is available as an option in NON-SMOKER$^\mathrm{WEB}$ since version 4.0w \cite{websmoker}. The new code SMARAGD \cite{smaragd,raureview,cyburt} will have an improved version of this as default and thus implicitly account for low NLD at the compound formation energy. Preliminary results with this modification are shown in Fig.\ \ref{fig:dc} (right). The code SMARAGD will also include a global DC treatment using an averaged DC model and energy-dependent spectroscopic factors \cite{bon07,raujpg,raureview}. This average DC approach aims at providing robust predictions despite of considerable differences between microscopic predictions \cite{rau98}. Preliminary results for this DC are shown in Fig.\ \ref{fig:dc} (right), too.

The final rate (or cross section) is the sum of the modified HF value and the DC one. Interestingly, for the isotopes shown here (except for $N=92$) this sum is approximated by the unmodified HF result within a factor of 10. This is in accordance with \cite{gordc} (see fig.\ 3 therein). This shows that it seems justified to use
unmodified HF rates as crude estimate of the total rates for exotic nuclei.

\end{document}